\documentclass[twocolumn]{aastex62}

\newcommand{\ms}[1]{$#1~{\rm m\,s^{-1}}$}

\newcommand{\vrot}{$v_{\rm rot}$}
\newcommand{\vkep}{$v_{\rm kep}$}
\newcommand{\dvrot}{$\delta v_{\rm rot}$}

\defcitealias{Isella_ea_2016}{I16}
\defcitealias{Liu_ea_2018}{L18}

\graphicspath{{./}{figures/}}

\received{2018 April 25}
\revised{2018 May 14}
\accepted{2018 May 16}
\submitjournal{ApJL}

\shorttitle{A Kinematical Detection of Two Jupiter Mass Planets}
\shortauthors{Teague et al.}

\begin{document}

\title{A Kinematical Detection of Two Embedded Jupiter Mass Planets in HD~163296}

\correspondingauthor{Richard Teague}
\email{rteague@umich.edu}

\author[0000-0002-0786-7307]{Richard Teague}
\affil{Department of Astronomy, University of Michigan, 311 West Hall, 1085 S. University Ave, Ann Arbor, MI 48109, USA}

\author[0000-0001-7258-770X]{Jaehan Bae}
\affil{Department of Terrestrial Magnetism, Carnegie Institution for Science, 5241 Broad Branch Road, NW, Washington, DC 20015, USA}

\author[0000-0003-4179-6394]{Edwin A. Bergin}
\affil{Department of Astronomy, University of Michigan, 311 West Hall, 1085 S. University Ave, Ann Arbor, MI 48109, USA}

\author[0000-0002-1899-8783]{Tilman Birnstiel}
\affil{University Observatory, Faculty of Physics, Ludwig-Maximilians-Universit\"{a}t M\"{u}nchen, Scheinerstr. 1, D-81679 Munich, Germany}

\author[0000-0002-9328-5652]{Daniel Foreman-Mackey}
\affil{Center for Computational Astrophysics, Flatiron Institute, 162 5th Ave, New York, NY 10010}

\begin{abstract}
We present the first kinematical detection of embedded protoplanets within a protoplanetary disk. Using archival ALMA observations of HD~163296, we demonstrate a new technique to measure the rotation curves of CO isotopologue emission to sub-percent precision relative to the Keplerian rotation. These rotation curves betray substantial deviations caused by local perturbations in the radial pressure gradient, likely driven by gaps carved in the gas surface density by Jupiter-mass planets. Comparison with hydrodynamic simulations shows excellent agreement with the gas rotation profile when the disk surface density is perturbed by two Jupiter mass planets at 83~au and 137~au. As the rotation of the gas is dependent on the pressure of the total gas component, this method provides a unique probe of the gas surface density profile without incurring significant uncertainties due to gas-to-dust ratios or local chemical abundances which plague other methods. Future analyses combining both methods promise to provide the most accurate and robust measures of embedded planetary mass. Furthermore, this method provides a unique opportunity to explore wide-separation planets beyond the mm~continuum edge and to trace the gas pressure profile essential in modelling grain evolution in disks.
\end{abstract}


\section{Introduction}
\label{sec:introduction}

Despite the detection of close to 4000 fully formed planets, there are only a handful of planets detected during their formative stages. Characterising the formation environment and witnessing the delivery of volatiles to the atmosphere of a young planet is an essential step in understanding the planet formation process. Towards this end, the Atacama Large Millimetre Array (ALMA) has revealed a variety of substructure in continuum emission from protoplanetary disks suggestive of planet-disk interactions sculpting the dust density distribution \citep{ALMA_ea_2015, Andrews_ea_2016, Isella_ea_2016, Zhang_ea_2016}.

Inferences of embedded planets and estimates of their mass are typically made from these observations by extrapolating a surface brightness profile of the thermal mm continuum to a gas surface density which can, in turn, be compared with either analytical prescriptions \citep[e.g.,][]{Duffell_2015} or directly to hydrodynamic simulations \citep{Fedele_ea_2018}. Similar perturbations observed in molecular line emission have been used as additional constraints on the gas surface density profile \citep{Isella_ea_2016, Fedele_ea_2017, Teague_ea_2017}.

It is well known from attempts to measure the total disk gas mass that the extrapolation of continuum and molecular line emission to a gas surface density is fraught with uncertainty \citep{Bergin_ea_2013, Miotello_ea_2017}. Ill-constrained gas-to-dust ratios, radially varying grain opacities, complex grain evolution dictated by a coupling of the solid particles to the gas \citep{Birnstiel_ea_2012} and local chemical and excitation effects \citep{Oberg_ea_2015, Cleeves_2016} all conspire to limit the accuracy of the recovered surface density profile.

It has been shown that a gap is not a unique signature of a planet. Massive planets are able to excite spiral waves which open up secondary and tertiary gaps \citep{Bae_ea_2017, Fedele_ea_2018}, while grain growth around ice lines \citep{Zhang_ea_2015} and the shepherding of dust by (magneto-)hydrodynamic instabilities have also been shown to produce ring like structure in the continuum \citep{Flock_ea_2015, Birnstiel_ea_2015, Okuzumi_ea_2016}. In total, these effects can result in a significant uncertainty in the derived planet mass and, in some cases, lead to an incorrect inference of a planet.

\begin{figure}
\centering
\includegraphics[]{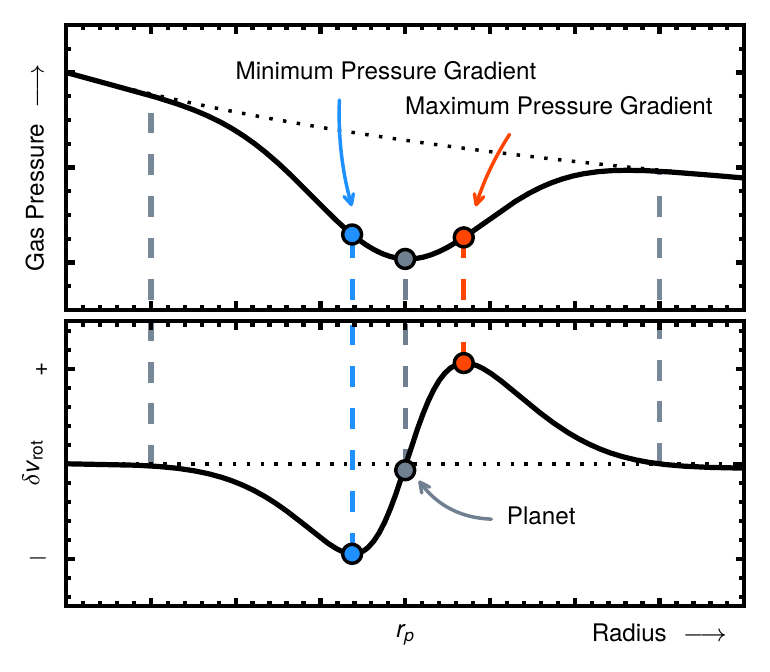}
\caption{A cartoon demonstrating the impact of a local gas pressure minimum (top) and the induced perturbations on the rotation velocity (bottom). \label{fig:cartoon}}
\end{figure}

An alternative approach is to use the rotation of the gas to probe the local pressure gradient. As the disk is in both radial and vertical hydrostatic equilibrium, the rotation velocity is given by

\begin{equation}
\frac{v_{\rm rot}^2}{r} = \frac{GM_{\star} r}{(r^2 + z^2)^{3/2}} + \frac{1}{\rho_{\rm gas}} \frac{\partial P}{\partial r},
\label{eq:rotation_full}
\end{equation}

\noindent where $M_{\star}$ is the mass of the star and $\partial P/\,\partial r$ is the radial pressure gradient \citep{Rosenfeld_ea_2013}. We have neglected the impact of self-gravity as this term requires knowledge of the disk mass. \citet{Rosenfeld_ea_2013} demonstrate that this term will introduce a slight hastening of the rotation at large radii for only the most massive disks and will not introduce small scale perturbations. When a planet perturbs the local gas density there will be a change in the local pressure gradient which manifests as a change in the rotation velocity as shown in Fig.~\ref{fig:cartoon} \citep[see also][]{Kanagawa_ea_2015}. As this method traces the pressure \emph{gradient} the deviations in rotation velocity, $\delta v_{\rm rot} = (v_{\rm rot} - v_{\rm kep}) / v_{\rm kep}$, will inform us on the \emph{shape} of the perturbation. By directly tracing the gas pressure, this technique is free from the uncertainties discussed above which traditional methods are prone to.

In this Letter we present a new method to precisely measure $v_{\rm rot}$ which is used to make the first kinematical detection of embedded planets in the disk of HD~163296, a A1 star at 101.5~pc \citep{Bailer-Jones_ea_2018}\footnote{This work was undertaken assuming the pre-Gaia distance of 122~pc \citep{vandenAncker_ea_1997} with all linear scales calculated under that assumption. In the following we do not rescale our models nor the results from previous work to allow for a more direct comparison.}. This source has previously been suggested to play host to multiple planets due to the substructure observed in its continuum and CO line emission \citep[hereafter \citetalias{Isella_ea_2016} and \citetalias{Liu_ea_2018}, respectively]{Isella_ea_2016, Liu_ea_2018}. 

In Section~\ref{sec:observations} we briefly describe the archival data used before presenting the method and results in Sections~\ref{sec:methodology} and \ref{sec:hydrosims}. We discuss the robustness of the method in Section~\ref{sec:discussion} and provide a summary in Section~\ref{sec:summary}. 

\section{Observations}
\label{sec:observations}

We use archival data of the $J = 2-1$ transitions of $^{12}$CO, $^{13}$CO and C$^{18}$O from the disk HD~163296 (2013.1.00601.S, PI Isella). These data were originally presented by \citetalias{Isella_ea_2016} who reported dips in the emission profile coincident with ringed substructure in the mm continuum. Modelling of this emission suggested the presence of three planets of 0.46, 0.46 and 0.58~$M_{\rm Jup}$ at 59, 105 and 160~au \citepalias{Liu_ea_2018}.

Our reduction closely follows the procedure outlined in \citetalias{Isella_ea_2016}. The data were initially calibrated with \texttt{CASA v4.4.0} before transferring to \texttt{CASA v4.7.2} for self-calibration and imaging. Phase self-calibration was performed on the continuum by combining the three continuum spectral windows before being applied to the whole dataset. No amplitude self-calibration was performed. Continuum emission was subtracted from the line data using the task \texttt{uvcontsub}, then the lines imaged using Briggs weighting and a robust value of 0.5 resulting in a beam size of $0.26\arcsec \times 0.18\arcsec$ for all three lines. The channel spacing was averaged down to \ms{50} in order to improve signal to noise. We refer the reader to \citetalias{Isella_ea_2016} for a discussion about the morphology of the emission.

\section{Precise Measurement of Rotation}
\label{sec:methodology}

\begin{figure*}
\centering
\includegraphics[width=\textwidth]{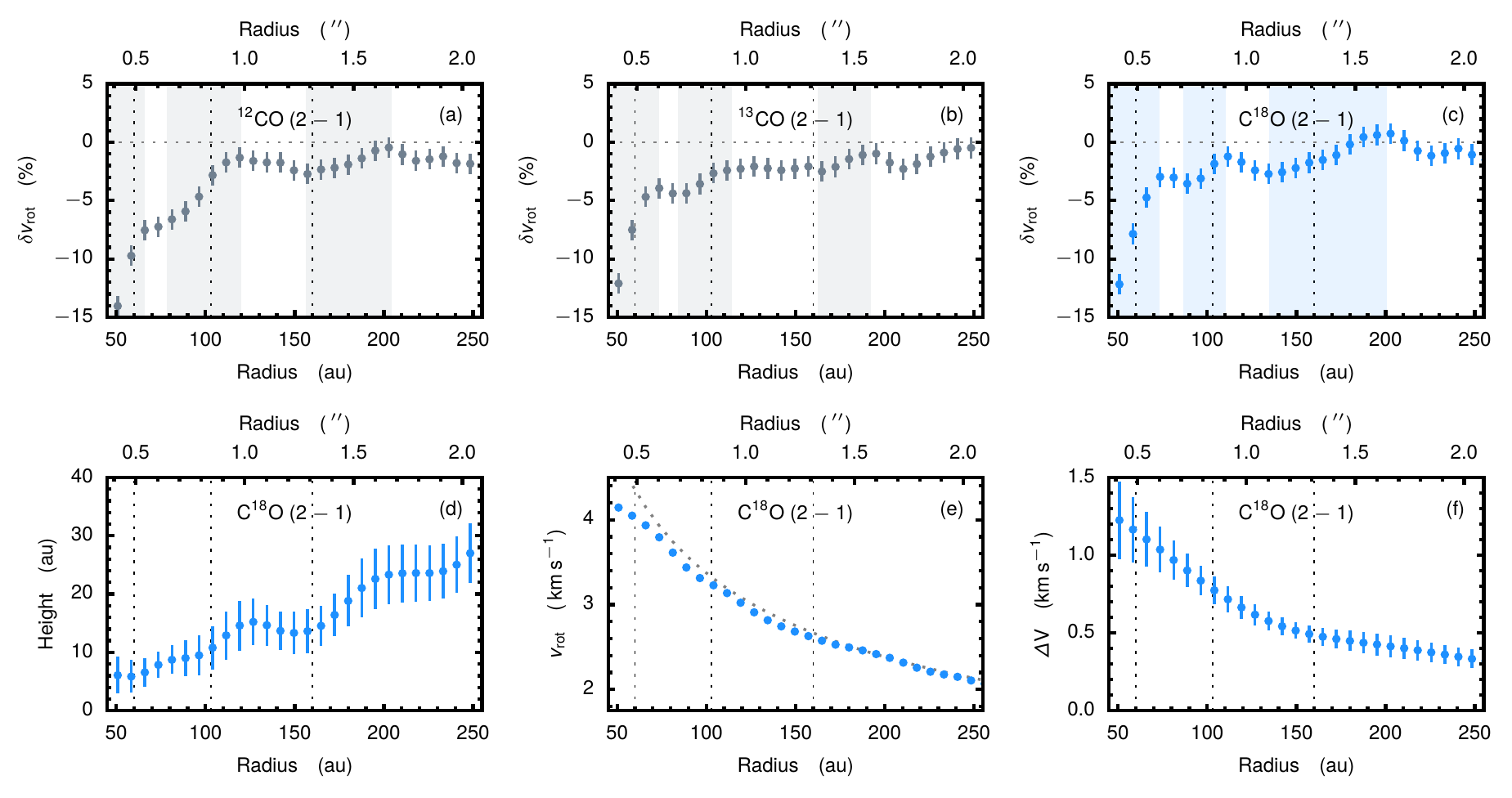}
\caption{Top row: \dvrot{} relative to the reference Keplerian curve for $^{12}$CO, $^{13}$CO and C$^{18}$O, left to right. Shaded regions highlight the characteristic \dvrot{} features associated with pressure minima. Bottom row: emission height, (d); \vrot{} compared to a Keplerian curve, (e) and linewidth, (f) for C$^{18}$O. Vertical lines show the centres of the gaps in the continuum emission. Error bars signify a $3\sigma$ uncertainty. \label{fig:observations}}
\end{figure*}

As the emission arises from a vertically elevated region with a lower limit set by the freeze-out of the CO molecules \citep{Rosenfeld_ea_2013, Schwarz_ea_2016}, we first measure the emission height to a) accurately deproject pixels into radial bins and b) account for any contributions to \dvrot{} from changes in $z$ (see Eqn.~\ref{eq:rotation_full}). For this we follow the method presented in \citet{Pinte_ea_2018}, using a position angle of 132\degr{} which we obtained from fitting a Keplerian rotation pattern to the first moment map and is consistent with previous determinations \citep{Flaherty_ea_2015, Flaherty_ea_2017}. The emission surface is modelled as a Gaussian Process (GP) which assumes that the observations are drawn from a smoothly varying function. This approach removes the need to assume an analytical form for the function, providing more flexibility in the model in addition to not requiring the binning of data \citep{Foreman-Mackey_ea_2017}. The resulting GP model for the C$^{18}$O emission is presented in Fig.~\ref{fig:observations}d showing slight depressions in the emission surface over the gap locations. This is indicative of a reduction in the local gas scale height or a perturbation in the CO column density.

To derive \vrot{}, we exploit the azimuthal symmetry of the system and build upon the spectral deprojection and azimuthally averaging technique used in previous studies of protoplanetary and debris disks \citep{Teague_ea_2016, Yen_ea_2016, Matra_ea_2017}. At a location $(r,\,\theta)$, where $r$ is the radius and $\theta$ is polar angle relative to the major axis of the disk, the centre of an emission line will be Doppler shifted by an amount $v_{\rm rot} \cdot \cos \theta$. Conversely, other properties of the line profile, such as amplitude and width vary only a function of radius and are constant around $\theta$. Thus, correcting for this offset allows for the lines to be stacked to provide a significant boost in the signal-to-noise ratio (SNR). Rather than assuming \vrot{} \emph{a priori} for this deprojection, we are able to infer its value from the data. Any error in the assumed \vrot{} value will result in slightly offset line profiles, resulting in a broadened stacked profile. We assert that the correct $v_{\rm rot}$ is the value which minimizes this width. 

This approach was validated on a suite of forward models with known \vrot{}. Although the chosen bin width is below the spatial resolution of the data, wider bins were found to sample a sufficiently large range of \vkep{} that would overwhelm any signal from changes in the pressure gradient. For a SNR of $\gtrsim 10$, as with the observations, we could recover \vrot{} accurately to a precision of \ms{2}.

To apply this to the observations we first binned the data into annuli roughly two pixels wide ($\sim 9$~au). To derive \vrot{} we use the \texttt{L-BFGS-B} method implemented in the \texttt{scipy.optimize} package. Panel (e) in Fig.~\ref{fig:observations} shows the derived \vrot{} profile for C$^{18}$O in blue. The uncertainties on individual bins are roughly \ms{2} between 50 and \ms{250}, while the uncertainty on the GP models, shown in the top row of Fig.~\ref{fig:observations}, is larger at \ms{8}. We find a larger uncertainty on the GP model than the individual bins because the extreme flexibility of the GP model makes it harder to distinguish between small scale fluctuations and noise, unlike when fitting an analytical function. In essence, precision in the resulting model is sacrificed to allow for a larger population of smooth models to be considered. Relative residuals from a Keplerian curve assuming $2.3~M_{\rm sun}$ and $i = 47.7\degr$, appropriate for HD~163296 \citep{Flaherty_ea_2015, Flaherty_ea_2017} are shown for the three transitions in the top row of Fig.~\ref{fig:observations}. Clear deviations are observed centred on the gaps in the continuum \citepalias[shown by the vertical dotted lines;][]{Isella_ea_2016}, consistent with the predicted profile shown in Fig.~\ref{fig:cartoon}. The significant deviation at smaller radii is an effect of the finite resolution of the observations. Testing with forward models show that this method is significantly biased within two beam FWHM from the disk centre, equal to $\approx 52$~au for the current observations.

\begin{figure}
\centering
\includegraphics[width=\columnwidth]{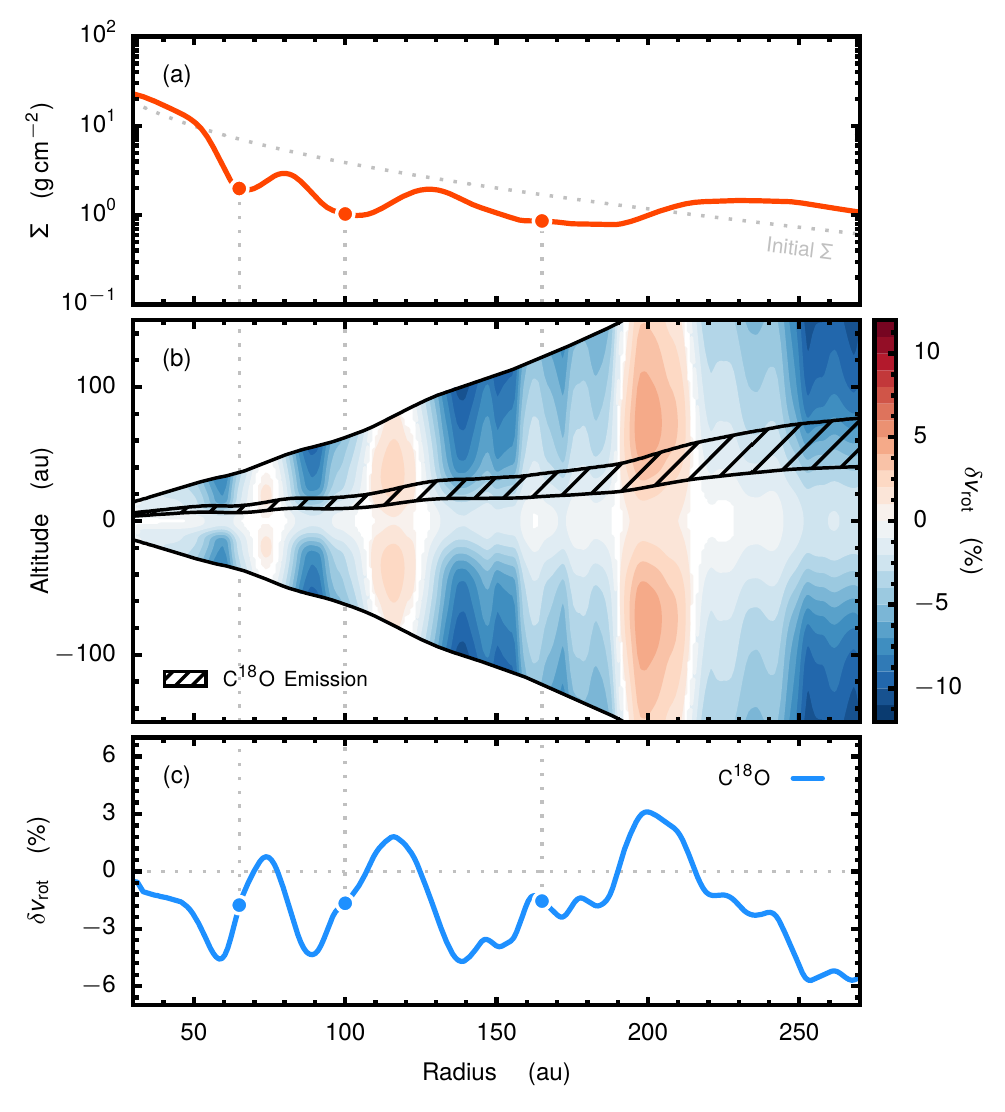}
\caption{Physical structure of the best-fit model. (a) Gas surface density profile with the planet locations shown by the dots. (b) Inflated physical structure used for the radiative transfer. The background colors show the deviation from cylindrical Keplerian rotation while the hatched region shows where the C$^{18}$O emission arises from, consistent with the measured height in Fig.~\ref{fig:observations}. (c) \dvrot{} profile from the C$^{18}$O emission layer account for local changes in gas density and temperature and height in the model. \label{fig:modelling}}
\end{figure}

\section{Hydrodynamic Models}
\label{sec:hydrosims}

In the following we focus on the C$^{18}$O emission as this arises from a region closer to the disk midplane so that 2D hydrodyanmic models are sufficient to model the velocity perturbations. Conversely $^{12}$CO and $^{13}$CO emission arises from much higher altitudes and thus requires the use of considerably more computationally expensive 3D hydrodynamic models. As such, the interpretation of these lines is left for future work focussing on the vertical structure of the perturbation.

Using Eqn.~\ref{eq:rotation_full} we are able to quantify the relative importance of the height, gas temperature and density. Although all three factors likely act in concert, isolating the main sources of perturbations helps in constraining the physical processes at play. Changes in the emission height can account for deviations in \vrot{} of $\lesssim 0.5\%$ at 165~au, however no significant change in $z$ is observed at 100~au or closer in. If the deviations were solely due temperature perturbations then we require changes in temperature of 45\% and 75\% at 100 and 165~au, respectively. These correspond to changes in linewidth of 19\% and 50\%, and changes in the integrated intensity of 48\% and 87\%. Such large temperature changes can be ruled out as the linewidth which shows deviations of $\lesssim 10\%$ across the gaps (Fig.~\ref{fig:observations}f) and the radial C$^{18}$O intensity profile \citepalias[Fig.~1 of][]{Isella_ea_2016} shows deviations of $\approx 50\%$ and $\lesssim 25\%$ at 100 and 165~au. This suggests that while changes in temperature are likely present, the dominant source of the perturbation is changes in the density profile, consistent with the previously proposed embedded planet scenario \citepalias{Isella_ea_2016, Liu_ea_2018}. Therefore in the following we focus on changes in the density structure and leave the effects of temperature for future work.

\begin{figure*}
\centering
\includegraphics[width=\textwidth]{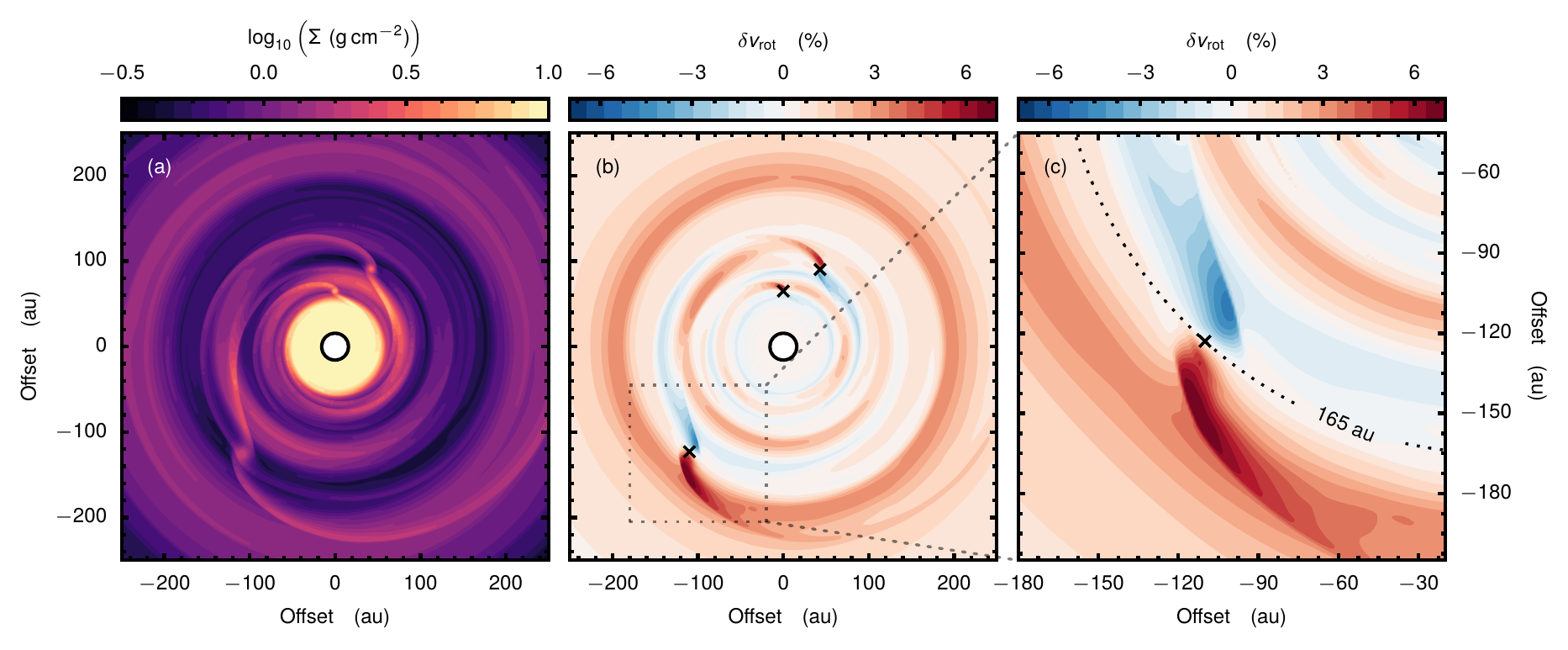}
\caption{(a) Surface density profile from the best-fit hydrodynamic simulation. (b) \dvrot{} due to changes in local pressure. (c) A zoom in of the \dvrot{} structure at the location of the outer planet. \label{fig:2dhydro}} 
\end{figure*}

To test the hypothesis of planetary causes for \dvrot{}, we ran hydrodynamical simulations of embedded protoplanets using \texttt{FARGO 3D} \citep{Benitez_ea_2016} including the orbital advection algorithm \texttt{FARGO} \citep{Masset_2000}. These used a polar grid with 1024 logarithmically spaced radial cells between 16 and 480~au and 1920 uniformly spaced azimuthal cells covering $2\pi$ radians. The underlying physical structure was taken from the best-fit model of \citet{Flaherty_ea_2017}.

We opted to run 2D simulations in order to efficiently explore the parameter space. An isothermal equation of state was assumed and a uniform viscosity of $\alpha = 10^{-3}$ was adopted, consistent with the constraints on the turbulence level in HD~163296 \citep{Flaherty_ea_2015, Flaherty_ea_2017}.  The calculations run for 1000 orbital times for on the outermost planet, corresponding to 4000 and 2100 for the innermost and middle planet, respectively. We found that in all simulations that the gas densities and velocities had reached steady state before the end of the simulation.

To compare these simulations to the observations we generate C$^{18}$O image cubes, simulate the observations and apply the methodology outline in Section~\ref{sec:methodology} to the simulated observations. This approach will fully account for changes in the emission height (and subsequent changes in the temperature) which can arise from perturbed surface densities. To generate the C$^{18}$O image cubes we inflated the azimuthally averaged surface density from the simulations assuming hydrostatic equilibrium. Above the 19~K isotherm and below a shielding column of $1.2 \times 10^{21}~{\rm H_2~cm^{-2}}$ \citep{Visser_ea_2009} a molecular abundance of $8.67\times10^{-8}$ was assumed for C$^{18}$O. Figure~\ref{fig:modelling} shows the azimuthally averaged physical structure of the best-fit model and the resulting \dvrot{} profile which is consistent with the observed profile in Fig.~\ref{fig:observations}. Figure~\ref{fig:2dhydro} shows the simulations before azimuthal averaging, showing the density structure in (a), \dvrot{} in (b), and a zoom in around the planet at 165~au. 

Radiative transfer was performed with the non-LTE code \texttt{LIME} \citep{Brinch_ea_2010} with image properties matching the observations. As we do not expect significant spatial filtering from the data \citepalias{Isella_ea_2016}, the images were convolved with a 2D Gaussian beam consistent with the C$^{18}$O observations to provide a fair comparison. We find an excellent match to the observations with a $1~M_{\rm Jup}$ planet at 100~au, and $1.3~M_{\rm Jup}$ planet at 165~au as shown in Fig.~\ref{fig:2dhydro}, however were unable to find a convincing fit to the feature at $< 70$~au. A planet of $0.6~M_{\rm Jup}$ at 65~au provided the best fit, however other potential scenarios are discussed in Section~\ref{sec:discussion}. We were able to constrain the mass of the two outer planets to within 50\% and their locations to within 10\%. Figure~\ref{fig:bestfit} compares the best-fit hydrodynamic model in blue with the observed \dvrot{} profile. The orange line shows the initial, unperturbed model demonstrating that a smooth model is a poor match to the data.

\section{Discussion}
\label{sec:discussion}

\begin{figure}
\includegraphics[width=\columnwidth]{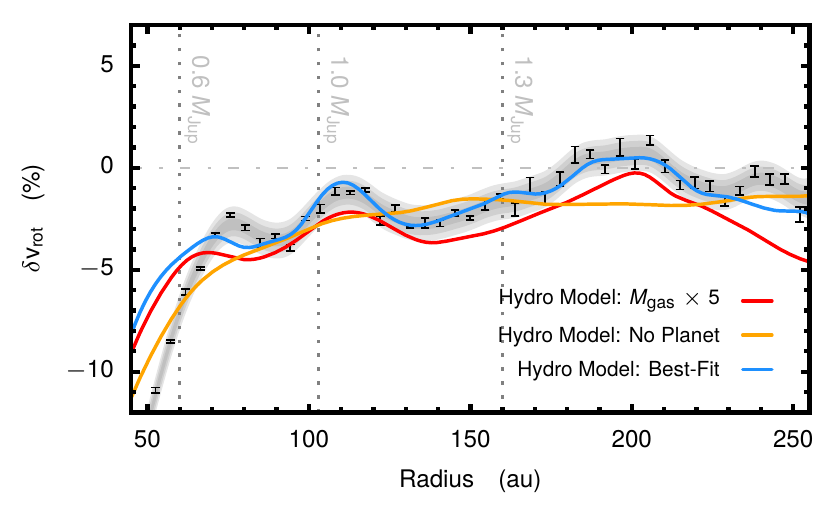}
\caption{Comparison of the simulations (three colored lines) with observations. The black error bars represent the individual annuli while the gray filled region represents the GP model. \label{fig:bestfit}}
\end{figure}

The results derived from this method are dependent on relative flux measurements rather than absolute intensities as in previous works. To demonstrate the robustness of this method to the model uncertainties, we present a model where we have increased the mass of the disk by a factor of five through a rescaling of the surface density (and thus an increase in C$^{18}$O column density). This model is plotted in red in Figure~\ref{fig:bestfit}. The derived \dvrot{} profile is minimally affected other than a slight shift to lower values as the emission arises from a higher (and therefore slower rotating) region. Conclusions based on the presence and mass of the planets are therefore insensitive to the disk mass and thus these constraints are exceptionally complimentary to current methods. 

For this pilot study we have not included any explicit heating or cooling across the gaps as this introduces additional uncertainty. The level of heating depends on the accretion assumed onto the planet which can vary by orders of magnitude for a specific planetary mass \cite{Mordasini_ea_2017}, while changes in the thermal structure due to changes in the dust distribution require computationally expensive modelling of the 2D grain distribution \citep{Teague_ea_2017, Facchini_ea_2017}. As we have not included cooling of the gas within the gap our planet masses should be seen as an upper limit and the inclusion of a decrease in temperature in the gaps could account for the difference in derived masses between this work and that in \citetalias{Liu_ea_2018}.

Although in no way optimized to reproduce emission profiles, our best fit model is able to broadly recover the observed deviations in the C$^{18}$O normalised integrated intensity profiles \citepalias[Fig.~1,][]{Isella_ea_2016}. Deriving a model consistent with both the velocity signatures and the radial emission profiles is left for future work. Such an approach will result in unparalleled constraints on the physical conditions around planet-opened gaps.

As with \citetalias{Isella_ea_2016}, no reasonable fit was found for the \dvrot{} perturbation at 80~au with a single planet, even the continuum ring is too wide to be well described by multiple embedded planets. An alternative solution for this perturbation is the pressure confinement of grains at 80~au due to the edge of the deadzone of the magneto-rotational instability \citep{Flock_ea_2015, Pinilla_ea_2016b}. Such pressure confinement would require a pressure maximum at the centre of the bright continuum ring at 80~au, and thus a local maximum in \dvrot{} slightly inwards of this location. Recovery of the rotation profile is limited here by spatial resolution and higher observations would be sufficient to see whether a local minimum in \dvrot{} could be resolved at $\approx 45$~au, thereby distinguishing between these scenarios.

As we have unparalleled constraints on the gas surface density profile and the size and shape of the perturbations, we are able to rule out other scenarios which may produce similar features in the gas surface density. For example, the vertical shear instability has been shown to also produce concentric rings and gaps with a width much narrower than the gas scale height \citep{Flock_ea_2017}. This would result in gaps far narrower than those required to match the \vrot{} observations at 100 and 165~au. Similarly, ambipolar diffusion driven self-organisation has been shown to perturb the surface density \citep{Bethune_ea_2017}, however results in rings of constant width across the disk, again inconsistent with the observations. Finally, reconnection of magnetic field lines can produce ringed sub-structure \citep{Suriano_ea_2018}, however the strength of this in the outer disk, $r \gtrsim 10$~au, has yet to be demonstrated. For the outer two perturbations, only the embedded planet scenario is able to succinctly account for all the observations. Nonetheless, without the detection of a point source, we cannot unambiguously dismiss scenarios without a planet. 

\section{Summary}
\label{sec:summary}

We have presented a new method which enables the direct measurement of the gas pressure profile. This allows for significantly tighter, and more accurate, constraints on the gas surface density profile than traditional methods. Furthermore, as this method is sensitive to the gap profile, it provides essential information about the gap width in the gas which is typically poorly constrained from brightness profiles.

Application of this method to CO isotopologue emission from HD~163296 revealed the predicted deviations for a significant perturbation in the local pressure. By accurately measuring the height of emission and using the linewidth as a proxy of local gas temperature we were able to isolate changes in the local gas density as the primary drivers of these perturbations for C$^{18}$O.

Comparisons with hydrodynamic models found excellent agreement with a $1~M_{\rm Jup}$ planet at 100~au and a $1.3~M_{\rm Jup}$ at 165~au (83~au and 137~au at the new distance of 101.5~pc) and allowed for a constraint on the mass and radius of 50\% and 10\%, respectively. A fit to the inner perturbation was less successful and requires high resolution observations to distinguish between scenarios.

This method represents a new approach to searching for planets still embedded in their parental protoplanetary disk. By tracing the total gas component directly via the pressure, this method is free from the numerous uncertainties involved with more traditional approaches of mapping flux measurements to gas surface densities.

\acknowledgments
We would like to thank the referee for their helpful comments leading to a much clearer presentation of the results. This paper makes use of the following ALMA data: JAO.ALMA\#2013.1.00601.S. ALMA is a partnership of European Southern Observatory (ESO) (representing its member states), National Science Foundation (USA), and National Institutes of Natural Sciences (Japan), together with National Research Council (Canada), National Science Council and Academia Sinica Institute of Astronomy and Astrophysics (Taiwan), and Korea Astronomy and Space Science Institute (Korea), in cooperation with Chile. The Joint ALMA Observatory is operated by ESO, Associated Universities, Inc/National Radio Astronomy Observatory (NRAO), and National Astronomical Observatory of Japan. The National Radio Astronomy Observatory is a facility of the National Science Foundation operated under cooperative agreement by Associated Universities, Inc. T.B. acknowledges funding from the European Research Council (ERC) under the European Union’s Horizon 2020 research and innovation programme under grant agreement No 714769. J.B. acknowledges support from NASA grant NNX17AE31G and computing resources provided by the NASA High-End Computing (HEC) Program through the NASA Advanced Supercomputing (NAS) Division at Ames Research Center.

\vspace{5mm}
\facilities{ALMA}
\software{\texttt{emcee} \citep{emcee}, \texttt{celerite} \citep{Foreman-Mackey_ea_2017}, \texttt{CASA} \citep{casa}, \texttt{LIME} \citep{Brinch_ea_2010}, \texttt{FARGO-3D} \citep{Benitez_ea_2016}, \texttt{scipy} \citep{scipy}}

\end{document}